\documentclass[10.5pt,a4paper,twoside]{article}
\usepackage{epsfig}
\usepackage{baltlat5}
\usepackage{wrapfig}
\usepackage[authoryear]{natbib}
\begin{document}
\ \
\vspace{-0.5mm}

\setcounter{page}{1}
\vspace{-2mm}

\titleb{A Possible Orbital Solution For The Triple Star WDS 18253+4846}

\begin{authorl}
\authorb{Bojan Novakovi{\' c}}{}
\end{authorl}

\begin{addressl}
\addressb{}{Astronomical Observatory, Volgina 7, 11160 Belgrade 74, Serbia}
\end{addressl}

\begin{summary}
 We present here the orbital elements for the pair $BC$ of visual binary
 star WDS 18253+4846 (ADS 11344 = HD 170109) and ephemerides for the
 period 2008-2014. As the orbital elements of pair $AB$ already existed
 we analyzed mass ratio between these two pairs and concluded that components
 $C$ is the biggest one in this triple system.

\end{summary}

% Use the same key words as in AJ, ApJ or A\&A


\sectionb{1}{INTRODUCTION}

Although we are used to thinking of stars coming as individuals,
this is not the norm. The evidence is that most stars that we see
in the sky are parts of binary or even multiple star systems
revolving around a common center of mass. The classifications
include visual binaries, eclipsing binaries, and spectroscopic
binaries. Some binary systems are sufficiently close to Earth or
their mutual distance is large and the stars are well enough
separated that we can resolve their images in a telescope and
track their motion over a period of time. We call such systems
visual binaries.

Relatively large number of visual binaries are included in the
Washington Double Star Catalog (WDS)\citep{mas06}, but only for
small fraction of them (about 2 \%) an orbital solution existed
\citep{harmas06}. In the case of the systems with three or more
components, which are in general more important, orbital solutions
for all components of the system are calculated only in several
cases.

\sectionb{2}{METHODS AND RESULTS}

We present here a short discussion on the orbits of star WDS
18253+4846 (HIP 90284 = ADS 11344 = HD 170109). This is a triple
star and orbital solution for the pair HU 66 $AB$ already existed
\citep{sey02}. Using a Sector Grid Search (SGS) method
\citep{nov07} we calculated the orbital elements for pair HU 66
$BC$.

As there are only 7 measurements of pair $BC$ in the WDS catalog
and this is not enough for a reliable orbit determination we
calculated additional 31 relative positions of pair $BC$ using the
measurements of pairs $AB$ and $AC$ which are made at same epoch.
This allowed us to calculate orbital elements for pair BC from 38
relative positions. We assigned the appropriate weights to all
measurements according to the weighting rules described in
\citet{har89,har01}. In order to make these data consistent the
weights of derived positions are reduced by factor 2.

The obtained values of the orbital elements and their formal
errors for the pair $BC$ as well as the ephemerides for period
2008-2014 are presented in Table 1.

Finally, from these two sets of the orbital elements for the pairs
AB and BC we calculated the ratio between the sums of masses:

\begin{center}
   $q$ = $\frac{M_{C}+M_{B}}{M_{A}+M_{B}}$.
\end{center}

The obtained value of $q\approx6$ indicates that components $C$ is
the biggest one in this triple system. So we concluded that B
orbiting around $C$ and $A$ orbiting around $B$. This is in a good
agreement with measurements available for the pair $AC$ and
provides an explanation for "strange" motion of component $C$
(i.e. direction of motion alternate between direct and
retrograde).

\begin{table}
\centering \caption{Orbital elements and ephemerides} \label{}
\begin{tabular}{ccc} \hline\hline
Orbital elements  & \ \ &Ephemerides (2008-2014) \\
(J2000)  & &$\theta[^o]$ \ \ \ \ \ $\rho[^{\prime\prime}]$\\
\hline
$P[yr]$ = 561.20 $\pm$ 28.74&    & 293.9 \ \ \ 0.718 \\
$T$ = 2090.38 $\pm$ 26.82&    & 294.4  \ \ \ 0.723 \\
$a[^{\prime\prime}]$ = 1.103 $\pm$ 0.151&    & 294.9 \ \ \ 0.728 \\
$e$ = 0.637 $\pm$ 0.092&    & 295.4  \ \ \ 0.733 \\
$i[^o]$ = 108.9 $\pm$ 1.6&    & 295.8  \ \ \ 0.739 \\
$\Omega[^o]$ = 14.4 $\pm$ 2.0&    & 296.3 \ \ \ 0.745 \\
$\omega[^o]$ = 92.7 $\pm$ 2.4&    & 296.8 \ \ \ 0.752 \\
\hline \hline
\end{tabular}
\end{table}

\vskip5mm

ACKNOWLEDGMENTS. This research have made use of the Washington
Double Star Catalog maintained at the U.S. Naval Observatory and
it has been supported by the Ministry of Science of the Republic
of Serbia (Project No 146004 "Dynamics of Celestial Bodies,
Systems and Populations").

%\endreferences

\end{document}